\documentclass[10pt]{iopart}

\usepackage{times}
\usepackage{soul}
\usepackage{url}
\usepackage{xcolor}     

\usepackage{hyperref}
\hypersetup{
    colorlinks = true,
    linkcolor = purple,
    urlcolor  = blue,
    citecolor = purple,
    anchorcolor = black,
}

\usepackage[utf8]{inputenc}
\usepackage{graphicx}
\usepackage{amsthm}
\usepackage{booktabs}
\usepackage{algorithm}
\usepackage{algorithmic}
\usepackage{graphicx} 
\usepackage{xcolor}

\usepackage{enumitem}
\setlist[itemize]{leftmargin=*}

\usepackage{natbib}

\usepackage{breakcites}

\begin{document}

\title{Investigating naturalistic hand movements by behavior mining in long-term video and neural recordings}

\author{Satpreet H. Singh$^1$, Steven M. Peterson$^{2,6}$, Rajesh P. N. Rao$^{3,4,5}$, and Bingni W. Brunton$^{2,5,6,\dagger}$}

\address{$^1$ Department of Electrical and Computer Engineering, University of Washington, Seattle, USA}
\address{$^2$ Department of Biology, University of Washington, Seattle, USA}
\address{$^3$ Paul G. Allen School of Computer Science and Engineering, University of Washington, Seattle, USA}
\address{$^4$ Center for Neurotechnology, Seattle, USA}
\address{$^5$ University of Washington Institute for Neuroengineering, Seattle, USA}
\address{$^6$ eScience Institute, University of Washington, Seattle, USA}
\address{$\dagger$Author to whom correspondence should be addressed.}
\ead{bbrunton@uw.edu}


{
\begin{abstract}

\textit{Objective}:
Recent technological advances in brain recording and artificial intelligence are propelling a new paradigm in neuroscience beyond the traditional controlled experiment.
Rather than focusing on cued, repeated trials, \emph{naturalistic neuroscience} studies neural processes underlying spontaneous behaviors performed in unconstrained settings.
However, analyzing such unstructured data lacking \textit{a~priori} experimental design remains a significant challenge, especially when the data is multi-modal and long-term.
Here we describe an automated approach for analyzing simultaneously recorded long-term, naturalistic electrocorticography (ECoG) and naturalistic behavior video data.

\textit{Approach}:
We take a behavior-first approach to analyzing the long-term recordings. 
Using a combination of computer vision, discrete latent-variable modeling, and string pattern-matching on the behavioral video data, we find and annotate spontaneous human upper-limb movement events.
We then demonstrate applications of these naturalistic behavior events, along with their associated neural recordings, for neural encoding and decoding.

\textit{Main results}:
We show results from our approach applied to data collected for 12 human subjects over 7--9 days for each subject.
Our pipeline discovers and annotates over 40,000 instances of naturalistic human upper-limb movement events in the behavioral videos. 
Analysis of the simultaneously recorded brain data reveals neural signatures of movement that corroborate prior findings from traditional controlled experiments.
We also prototype a decoder for a movement initiation detection task to demonstrate the efficacy of our pipeline as a source of training data for brain-computer interfacing applications. 

\textit{Significance}:
Our work addresses the unique data analysis challenges in studying naturalistic human behaviors, and contributes methods that may generalize to other neural recording modalities beyond ECoG.
We publicly release our curated dataset, providing a resource to study naturalistic neural and behavioral variability at a scale not previously available.
\end{abstract}
}

%
\vspace{2pc}
%
%
\maketitle
%
\ioptwocol

\section{Introduction}
Neuroscience has long been interested in understanding brain activity associated with spontaneous behaviors in freely behaving subjects. 
Even so, hypotheses regarding brain function have typically been tested using carefully designed, well-controlled experimental tasks, where timing of cues, stimuli, and behavioral responses are known precisely. 
Fortunately, recent technological advances have enabled us to study increasingly naturalistic and longer brain recordings, giving rise to a new paradigm called ``naturalistic neuroscience” \citep{nastase2020keep, huk_beyond_2018, gabriel_neural_2019-1, markowitz_striatum_2018, wang_unsupervised_2016} where neural computations associated with such spontaneous behaviors are studied. 
Understanding such unstructured, long-term, and multi-modal data poses a substantial analytic challenge, due in part to the lack of \textit{a~priori} experimental design and the difficulty of isolating interpretable behavioral events. 

\begin{figure*}[bthp!]
    \centering
    \includegraphics[width=1.0\linewidth]{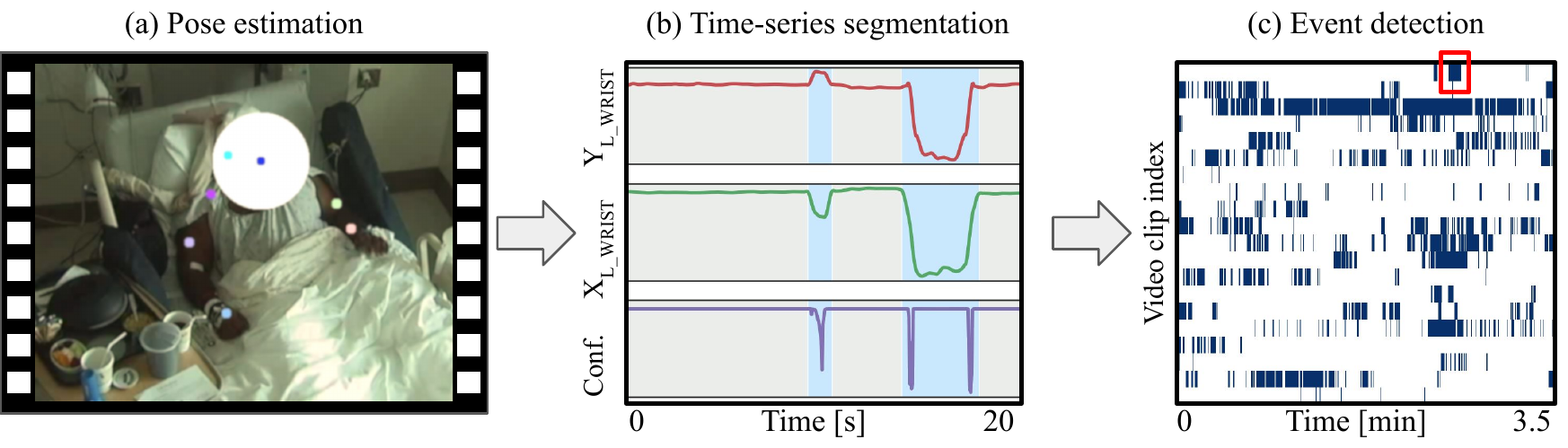}
    \caption{ 
    Pipeline for behavioral video data processing.
     (a) Video frame showing estimated pose keypoints (colored dots) on human subject.  
     (b) Autoregressive hidden Markov model (Section \ref{text:tss}) robustly segmented pose trajectory into \textit{rest} (shaded grey) and \textit{move} (shaded light blue) states. 
     (c) Raster plot of pose states (\textit{move} in dark blue, \textit{rest} in white) for several video-clips for pattern matching at scale. Red box depicts one \textit{movement initiation} event matching a pattern of 15 contiguous \textit{rest} states (0.5s) followed by 15 contiguous \textit{move} states (0.5s). 
    }
    \label{fig_pipeline}
\end{figure*}

\subsection{Related work}
Our work is related to several areas of active research in neuroscience, neuroengineering, and neuroethology that integrate techniques from machine-learning, computer-vision, and statistical modeling. 
Many recent methodological innovations have addressed the automated analysis of \textit{non-human} animal behavior \citep{batty2019behavenet, pereira2019fast, nassar2018tree, mathis_markerless_2018, johnson2016nips, wiltschko_mapping_2015} (see also \citet{mathis2020deep} for a recent survey and \citet{anderson_toward_2014} for a perspective on this emerging area).    
A typical non-human naturalistic neuroscience experiment \citep{johnson2020probabilistic, markowitz_striatum_2018, berman_measuring_2018} first collects simultaneously recorded behavioral video and neural activity data from one or more freely behaving subjects in an uncontrolled but sufficiently confined environment. 
Next, the video recordings are processed through an extensive pipeline consisting of steps such as: 
segmenting the subject(s) from the background, 
transforming subject pose to common coordinates using affine transformations, 
estimating pose of body-parts across frames, 
and higher-level operations such as classifying pose or segmenting pose into actions.
Combined with the simultaneously recorded neural data, such naturalistic behavior data are being used to shed light on previously intractable questions in behavioral neuroscience, often at unprecedented scale.

Human action-recognition methods from mainstream computer vision \citep{ramasamy2018recent} are relevant but not directly applicable to the needs of naturalistic human neuroscience. 
Traditionally, action-recognition research has concerned itself with discriminating activities at a coarse level, such as sitting vs. walking \citep{ghorbani2020movi}, and has often assumed the availability of a large corpus of labeled training data.
In contrast, to study the kinds of behaviors that interest neuroscientists and neuroengineers, we seek to localize fine-grained movements to sub-second temporal resolution, and ideally use the fewest behavioral labels possible \citep{seethapathi2019movement}. 
Lastly, since it is not known which behaviors or behavioral characteristics will elicit neural responses worth studying further, a queryable representation that supports the flexibility to study several kinds of behaviors is desirable. 
Recent work by \citep{fu_rekall_2019} develops such representations for semi-automated exploration of scenes in general videos.

Our work is most closely related to recent work in human naturalistic neuroscience that combine computer vision with opportunistic clinical brain recordings, including \citet{wang_unsupervised_2016}, \citet{alasfour2019coarse}, and \citet{chambers2019computer}. 
In particular, we build on the work of \citet{wang2018ajile} using similar video data and estimate the pose of human upper-body keypoints using neural-networks.
\cite{gabriel_neural_2019-1} use optical flow and image partitioning to detect coarse limb movements from video taken in a clinical setting similar to ours and develop neural decoders for detecting these movements from brain data. 
Compared to \citet{wang2018ajile}, who use a moving window heuristic on pose estimates to detect movements, we take a more principled approach to modeling the pose data.
This allows us to localize movement events with finer temporal-resolution and characterize entire movement trajectories, which in turn enables novel applications described later in the paper.
{
We also use newer, more efficient computer vision methods \citep{mathis2020deep, nath2019using} that allow us to process data at a scale that exceeds all of the aforementioned studies taken together in the number of subjects and duration of recordings analyzed.
Finally, we focus on curating, characterizing, and making our dataset available to the research community to foster further research and development in this area.
} 

\subsection{Our approach}
We present a scalable behavior-mining approach to analyze simultaneously recorded naturalistic brain and behavior data, obtained opportunistically from human subjects undergoing long-term clinical monitoring prior to epilepsy surgery. 
Our video processing pipeline (Figure~\ref{fig_pipeline}) first estimates the locations of keypoints (e.g. wrists and elbows) on the upper-body using a neural network trained on each subject~\citep{mathis_markerless_2018}. 
We then segment the trajectory of each keypoint in time using discrete latent-variable models, building a discrete representation of pose dynamics. 
Interestingly, having a discrete, sequential representation of upper-limb pose simplifies the problem of detecting behavioral \textit{events} to pattern-matching on strings. 
Using regular-expressions corresponding to patterns of interest, we discover thousands of interpretable events per subject---an order of magnitude more observations than in a typical controlled human experiment.
To study the rich naturalistic variability associated with these events, we also extract metadata including movement angle, magnitude, and duration. 

Next, we explore the use of these behavioral events for neuroscience and neuroengineering applications by analyzing the simultaneously recorded brain data. 
Event-averaged spectrograms associated with our naturalistic human upper-limb movement initiation events corroborate and strengthen previous findings from controlled experiments \citep{miller_spectral_2007} (see also \citep{peterson2020behavioral}).
Preliminary investigations also suggest that our workflow could produce data useful for training brain-computer interface (BCI) decoders; due to the use of larger sample sizes of training data representative of naturalistic variability, such decoders may perform more robustly in real-world deployments.   

Our key contributions in this paper are as follows.
First, we present a highly automated, novel workflow for analyzing simultaneously recorded naturalistic long-term human brain and behavioral video data. 
Second, we develop a domain-relevant, robust, temporally precise, and queryable representation of human upper-limb pose.
Third, to showcase our workflow, we demonstrate example applications in neuroscience and neuroengineering, suggesting that our approach and results are of broad interest.
Finally, to support open science and facilitate further research in this area, we release our curated dataset consisting of annotated naturalistic events and associated neural recordings.

\section{Dataset}
\subsection{Human subjects and data collection}
Our dataset consists of human intracranial electrocorticography (ECoG) \citep{parvizi2018human} neural recordings and simultaneously recorded behavioral video recordings, obtained opportunistically from 12 patients with epilepsy for the duration of each patient's long-term (7--9 days) continuous clinical observation. 
The University of Washington Institutional Review Board for the protection of human subjects approved our study and all patients provided their informed written consent.
Patient behavior was continuously recorded by a wall-mounted camera (RGB/infrared, 640 $\times$ 480 pixels) for real-time monitoring by an around-the-clock clinical team, except during intermittent equipment servicing or private times when the camera was switched off or turned away.
Patients were observed performing their daily activities (including talking, eating, watching TV, using a computer or phone, sleeping, receiving clinical care etc.) from the hospital bed while being tethered to a brain-recording interface.

Each patient had about 90 electrodes implanted under the skull and dura, directly on their brain surface, including either an \mbox{8 $\times$ 8} grid or two \mbox{8 $\times$ 4} grids of electrodes. 
These electrodes were placed to monitor a subset of cortical regions, predominantly in one brain hemisphere, as determined by individual clinical need (right hemisphere for 5 and left hemisphere for 7 patients). 
Grid electrodes are spaced 1 cm apart center-to-center.
The sampling rates for video and ECoG recordings were 30 fps and 1000 Hz respectively. 
Together, the ECoG and video (approximately 18 million frames for a week) totaled about 250 GB of data per patient. 

\subsection{ECoG and video data preprocessing}
We applied standard pre-processing steps to the ECoG data \citep{peterson2020behavioral, gabriel_neural_2019-1, miller2019library, miller_spectral_2007, schalk2007decoding}, including down-sampling to 500Hz, 60-Hz line noise removal, large-amplitude artifact removal, and  median-centering using a common reference across all electrodes.
All electrode positions were localized and converted to Montreal Neurological Institute and Hospital (MNI) coordinates using the Fieldtrip toolbox \citep{fieldtrip_ref, stolk2018} in MATLAB.
To aid interpretability, we restricted our analysis in this paper to the 64 grid electrodes covering one hemisphere per subject.
We excluded electrodes with recording issues, such as persistent presence of artifacts. 
However, data for all available electrodes are provided in the publicly released dataset accompanying this paper.
In most cases, we also did not analyze the neurally and behaviorally atypical data from the first two days of a patient's hospital stay, since patients were usually heavily medicated during this time while recovering from electrode implantation surgery. 

Before processing the video data, we manually inspected and annotated it at a coarse (every 3 minutes or so) level of granularity to create an \textit{omit-list} that was excluded from further processing.
The omit-list included long time-spans of sleep, times when a clinical or research team was actively working with the subject, private times, and times when applying computer-vision algorithms was impossible due to poor lighting conditions or severe occlusion of the subject's body. 
Almost all camera movement occurs around times when the clinical team is actively working with the patient. 
Removing these times results in a mostly steady recording configuration as seen in Figure \ref{fig_pipeline}(a).
We also labeled and excluded times when the clinical team had placed seizure restraints on the subjects' hands, since these limited mobility and gave rise to unnatural movements. 
Completing these manual annotations took about 6--12 hours per subject, depending on their clinical treatment regime, activity and sleep schedule, and length of hospital stay.
When analysing ECoG and video data together, the two data-streams were synchronized using metadata extracted using the equipment manufacturer's (Natus Medical Incorporated) software.

\section{Methods}

\subsection{Precision extraction of behavioral events at scale}
We developed and validated a pipeline to extract temporally precise, interpretable movement events, by processing the video data through pose-estimation, pose time-series segmentation, event detection and finally, event metadata extraction (Figure~\ref{fig_pipeline}). 

\subsubsection{Markerless pose estimation} \label{text:estimation}
To extract a subject's pose from raw video, we trained a state-of-the-art markerless pose estimation tool \citep{mathis_markerless_2018} known for its speed and data efficiency \citep{nath2019using}. 
For training data, we manually annotated around 1000 frames per subject chosen randomly from the entire duration of a subject's video data, preferentially sampling active, daytime hours over times when the subject was asleep. 
For each frame, we annotated up to 9 keypoints on the subject's body whenever visible \citep{PoseletsICCV09}. 
These keypoints were the nose, both wrists, elbows, shoulders, and ears (Figure~\ref{fig_pipeline}a).

During prediction, the pose-estimation tool produced the $(x, y)$ coordinates and a \textit{confidence} estimate between $[0, 1]$ for each keypoint per frame (Figure~\ref{fig_pipeline}b).
To quantify the performance of keypoint tracking, we estimated the pixel-wise RMS error to be 1.54 $\pm$ 0.13 s.d. pixels on the training data and 5.97 $\pm$ 1.96 s.d. pixels on the holdout data, both averaged across 12 subjects; 
holdout data is 5$\%$ of the manual annotations, excluding points below confidence threshold of $0.1$.
As an approximate scaling, $12$ pixels in the video span about $4$ cm in physical units, which is about the width of a human wrist.
We estimated this scale by comparing standard human measurements \citep{mcdowell2009anthropometric} with median distance between shoulder keypoints (in pixels) at movement onset for a few subjects.
On average, estimating pose for the entire duration of a subject's video took $400$ GPU-hours per subject using \texttt{AWS p2.16xlarge NVIDIA K80} GPUs. 
We denoised the estimated pose trajectories by median filtering (window length $11$ frames) and smoothing (window length $11$, $2$nd order Savitzky-Golay \citep{schafer2011savitzky} filter).

\subsubsection{Segmentation of pose trajectories} \label{text:tss}
Next, we segmented the pose time-series into discrete, interpretable states while preserving the temporal precision of the keypoint tracking.
We applied a first-order autoregressive hidden semi-Markov model (ARHSMM) \citep{murphymlpp} with two latent states to each keypoint's time-series (Figure~\ref{fig_pipeline}b for left-wrist). 
This model converts each keypoint's continuous pose dynamics into discretized dynamics consisting of \textit{rest} and \textit{move} \textbf{states}.
Using a semi-Markov, rather than a Markov model, accounts for the bias that limbs tend to be at \textit{rest} most of the time and mitigates unnecessary switching between latent states.
Similar to \citep{wiltschko_mapping_2015}, we fit the ARHSMM using the \texttt{pyhsmm-autoregressive} package in Python \citep{johnson2013hdphsmm}.
The resulting states are at video frame-rate resolution and the segmentation is relatively robust to variation in lighting, camera angle, and level of activity in the video.

\begin{figure}
    \centering
    \includegraphics[width=1.0\linewidth]{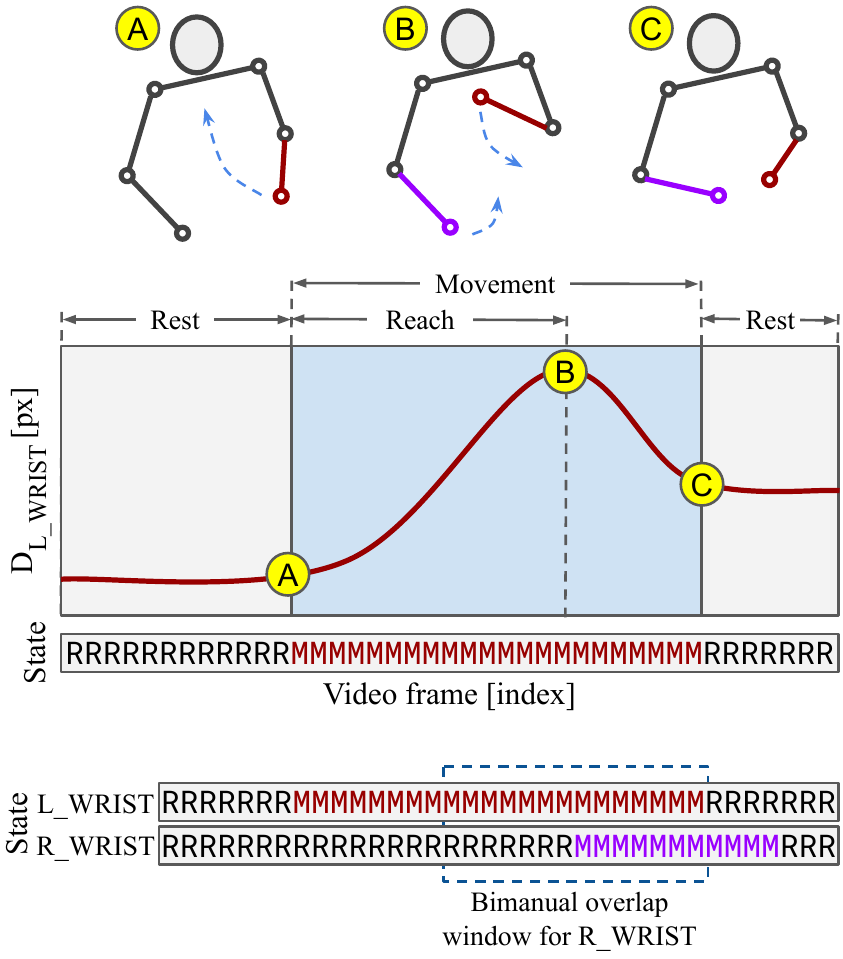}
    \caption{ Schematic of metadata extraction:
    [Top] Cartoon showing left-wrist movement at time of (A) movement initiation (B) maximum displacement from initiation (\textit{reach}) and (C) movement end. 
    [Middle] Time course of left-wrist radial distance [px: pixels], 
    and discretized state (R: \textit{rest}, M: \textit{move}).  
    Extracted movement metadata include duration, start and end coordinates, among others (Sec. \ref{text:metadata}).
    {
    [Bottom] Discretized state sequence for both left and right-wrists, now showing movement initiation in the right-wrist while the left-wrist is still in motion. 
    The dashed line shows the window over which \textit{bimanual overlap} metadata is calculated, corresponding to the number of frames for which the opposing (left) wrist is in motion over that duration. 
    }
    }
    \label{fig_cartoonbimanual}
\end{figure}

\subsubsection{Behavioral event mining} \label{text:mining}
Discretizing the pose trajectories facilitates the description of scientifically interesting behaviors performed spontaneously by the subject, even though they vary greatly in duration.
Specifically, the task of finding different types of behavioral \textbf{events} thus reduces to string pattern matching on the discretized dynamics.
For the behaviors we explore in the rest of this paper, we looked for \textit{movement initiation} events by matching a pattern of $15$ consecutive \textit{rest} states ($0.5$s), followed by at least $15$ consecutive \textit{move} states ($0.5$s).
Similarly, \textit{no-movement} events are state sequences of $90$ \textit{rest} states ($3.0$s) across both wrists and the nose.
To create our database of wrist movements, we use regular expressions to quickly find thousands of non-overlapping instances of such patterns in the discretized pose dynamics for each subject.

{
Parameters for smoothing, hyperparameters for the ARHSMM segmentation model, and the choice of regular expressions for event detection, were picked empirically by assessing performance on pose time-series derived from a small representative set of subject videos.
We confirmed that the temporal accuracy of event boundaries matched our expectations by manually inspecting a few dozen random events of each movement type for each subject. 
}

\subsubsection{Event metadata extraction} \label{text:metadata}
For each detected movement event, we extracted several metadata features from the continuous pose-dynamics associated with the movement. 
These include movement-associated metadata (Figure~\ref{fig_cartoonbimanual}) like the $(x,y)$ \textit{coordinates} of the keypoint at the start and end of the event,  \textit{duration} of the entire movement (up to next rest state), and \textit{rest duration} before and after movement.

Observed naturalistic hand movements often consisted of a hand reaching out, touching, or grabbing an object, then bringing the hand back to the body.  
Therefore, we defined the \textit{reach} of a wrist movement to be its maximum radial displacement during the course of the event, as calculated from its location at the start of the event. 
We extracted the \textit{magnitude}, \textit{angle}, and \textit{duration} for each reach. 

To measure the \textit{shape} of a movement, we fit $1^{st}$, $2^{nd}$ and $3^{rd}$-degree polynomials to a keypoint's displacement trajectory. 
Differences between the quality of the fit (as measured by $R^2$) to each polynomial type provide a rough measure of the ``curviness" of the movement trajectory. 
We also estimated a movement's onset and offset \textit{speeds}, by calculating the keypoint's displacement change within short time windows around the start and end of the movement. 

Since people often move both hands at the same time (i.e. ``bimanually"), we augmented each movement event with metadata about the opposing wrist's movement, if any (Figure~\ref{fig_cartoonbimanual}). 
By juxtaposing the discrete state sequence of both wrists, we calculated when the opposing hand starts to move (\textit{lead/lag time difference}) and how long this movement overlaps with that of the primary hand (\textit{overlap duration}).  

False positives in event discovery were still present in the data at this stage due to pose estimation failures and unusual pose states.
To compensate for failures in 2D pose estimation, we calculated movement-weighted \textit{confidence scores} for each event and removed those below a manually determined threshold. 
To eliminate outlier pose states, we calculated mean \textit{distance} and mean \textit{angle} between \textit{shoulder} keypoints, then removed events from the top and bottom 5 percentiles of these quantities.

\begin{figure*}[tb!]
    \centering
    \includegraphics[width=1.0\linewidth]{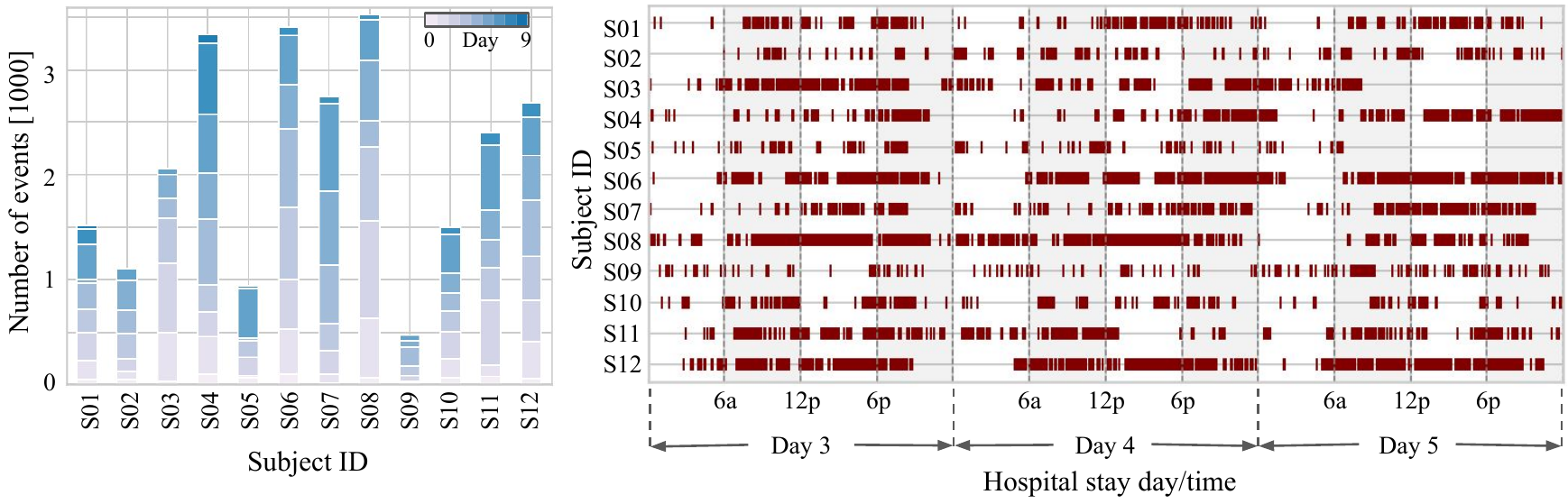}
    \caption{ 
    [Left] Number of right-wrist movement initiation events discovered per day for each of 12 subjects, totaling 475 to 3526 events per subject across their entire duration of clinical observation 
    (268 $\pm$ 123 s.d. per day).
    [Right] Raster plot of right-wrist movement initiation event occurrences showing bursts of activity interspersed with periods of rest or omit-listed (Section \ref{text:estimation}) periods.
    See Figure A\ref{fig_rastercounts_l} for equivalent plots for left-wrist.
    }
    \label{fig_rastercounts_r}
\end{figure*}

\begin{figure}
    \centering
    \includegraphics[width=1.0\linewidth]{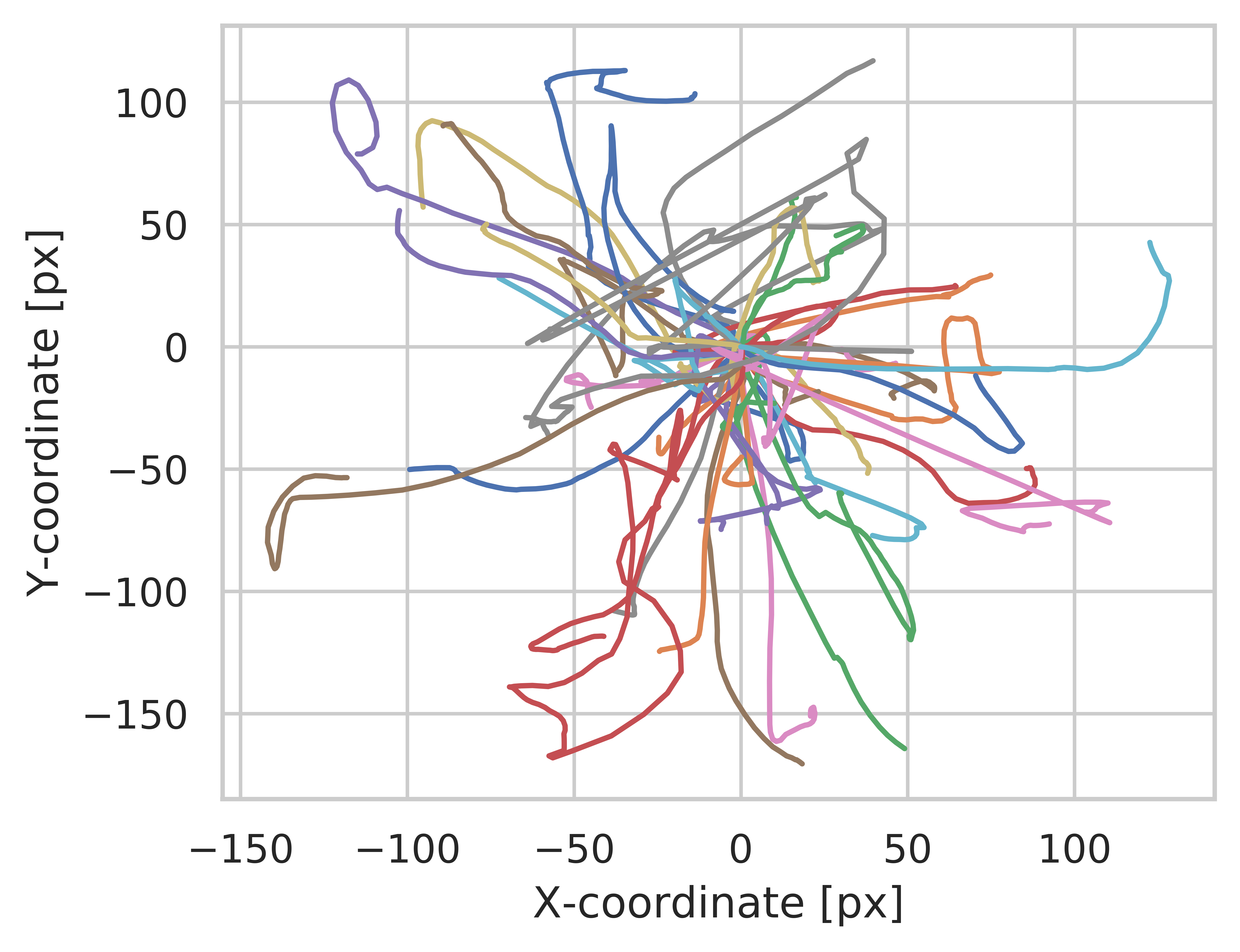}
    \caption{
    {
    A sample of 50 typical right-wrist trajectories (px: pixels; translated to start at origin) showing diversity of naturalistic reach movements for a single subject (S10).
    Different colors represent different individual trajectories.
    Note the large variability in the movements, compared to what is normally captured by controlled experiments.}
    }
    \label{fig_trajectories}
\end{figure}

\begin{figure*}[t]
    \centering
    \includegraphics[width=1.0\linewidth]{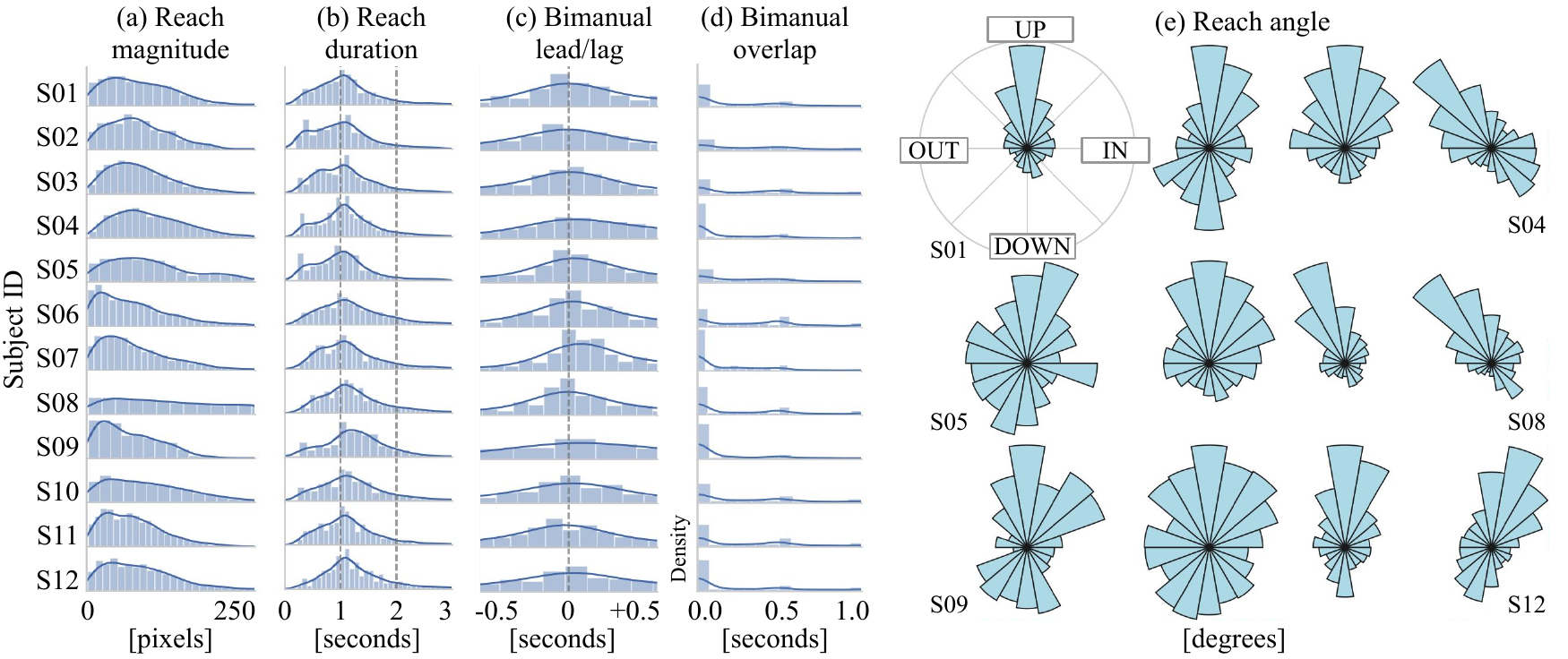}
    \caption{ 
    Histograms of right-wrist movement initiation event metadata per subject for their entire duration of clinical observation: 
    (a) Reach magnitude shows a dominance of small movements, 
    (b) Reach durations tended to be concentrated around $\approx$1s,
    (c) When both hands moved together (``bimanually"), they tended to start at about the same time.
    (d) Duration of time left hand was moving during a $\pm$0.5s window around time of right hand movement initiation.
    (e) 
    Polar histograms show that many subjects primarily made upward-downward reaches.
    However, reaches in almost every other direction were also observed.
    See Figure A\ref{fig_metadata_l} for equivalent plots for left-wrist.
    }
    \label{fig_metadata_r}
\end{figure*}

\subsection{Neural correlates of movement} \label{text:encoding}
A core scientific question in systems neuroscience is how behaviors are encoded by the coordinated activation of brain regions.
To examine the neural correlates of naturalistic movement initiation, we performed a time-frequency (TF) analysis of the neural recordings \citep{cohen_analyzing_2014} by averaging event-locked spectrograms for each subject, using hundreds of movement initiation events chosen to match movement statistics (reach magnitude, onset velocity, and shape) of a previous controlled experimental study \citep{miller_spectral_2007}. 
Using the aforementioned metadata to guide our search, we selected up to 200 events per day over 5 days for each of 12 subjects, and then further inspected the video for each event to remove any false positives (17.8$\%$ mean $\pm$ 9.9$\%$ s.d. events).

\subsection{Decoding naturalistic movement initiation} \label{text:decoding}
A grand challenge in neuroengineering is development of BCIs that can be used to predict spontaneous activity and intentions outside the lab, in everyday settings \citep{shanechi2019brain, smalley2019business, shanechi2018brain, warren2016recording, shenoy2012neural}.
Here we performed a preliminary study leveraging our pipeline as a source of training data for a BCI decoder that detects wrist movement initiation events. 
Specifically, we trained separate classifiers, tailored to each subject, to discriminate between movement initiation events and no-movement epochs for each wrist using only features derived from the ECoG neural recordings.


Our decoder uses the Random Forest (RF) algorithm \citep{Breiman:2001:RF:570181.570182, murphymlpp}, which is typically considered one of the best off-the-shelf classification algorithms for small/medium sized datasets \citep{hastie2009elements}. 
We used ECoG data $0.5$s before to $0.5$s after each event to compute TF spectrograms
at each of the 64 grid electrodes and used the flattened vector of TF bins as features for the classifier (TF bins were 200ms $\times$ 5Hz resolution, truncated at 150 Hz; approximately 9000 features total).

Given that the brain's response can drift over the course of days \citep{farshchian2018adversarial, klosterman2016day}, a reduced subset of events from 3 consecutive days (typically days 3 through 5 of clinical monitoring) were used.
We used events from the last day as the test set.
To eliminate the confound of movement initiation in the opposing wrist, we further filtered events to exclude those with significant movement ($\geq 0.2$ seconds) in the opposing wrist within the $\pm 0.5$s window used for ECoG data.
Positive (movement initiation) and negative (no-movement) examples were balanced by down-sampling negative examples. 
This balancing eliminated bias in the training set and set up a baseline performance of $50 \%$ accuracy for test set performance.
Training and test supports were 633 $\pm$ 417 s.d. and 331 $\pm$ 203 s.d. examples, respectively.
We tuned the RF using a 20-trial randomized search over two hyperparameters: number of trees (range: $[50, 250]$) and maximum tree-depth (range: $[3, 15]$). 
For each set of hyperparameters, 5-fold cross-validation holdout accuracy was used to measure performance. 
Final performance reported is from training using best hyperparameters and corresponds to classifier accuracy on events from the withheld test day.

\section{Results}

\subsection{Characterizing naturalistic events} \label{text:analytics}
Our pipeline extracted 959 to 6745 individual wrist movement events per subject (487 $\pm$ 215 per day) across 12 subjects (Figures \ref{fig_rastercounts_r} and A\ref{fig_rastercounts_l}). 
We found large variability between subjects in the number of events discovered, which we attribute to inter-subject differences in cycles of sleep and wakeful activity and clinical treatment regimes (Figure~\ref{fig_rastercounts_r} and A\ref{fig_rastercounts_l}).  
We also observed rich within-subject variability in the event metadata (Figures \ref{fig_trajectories}, \ref{fig_metadata_r}, and A\ref{fig_metadata_l}), which further differentiates our dataset from those collected in controlled experiments.    
Since our subjects received no instructions for when and how to move, we expect the observed movement statistics to closely reflect the natural statistics of human upper-limb movements while seated.

\begin{figure}[tb!]
    \centering
    \includegraphics[width=1.0\linewidth]{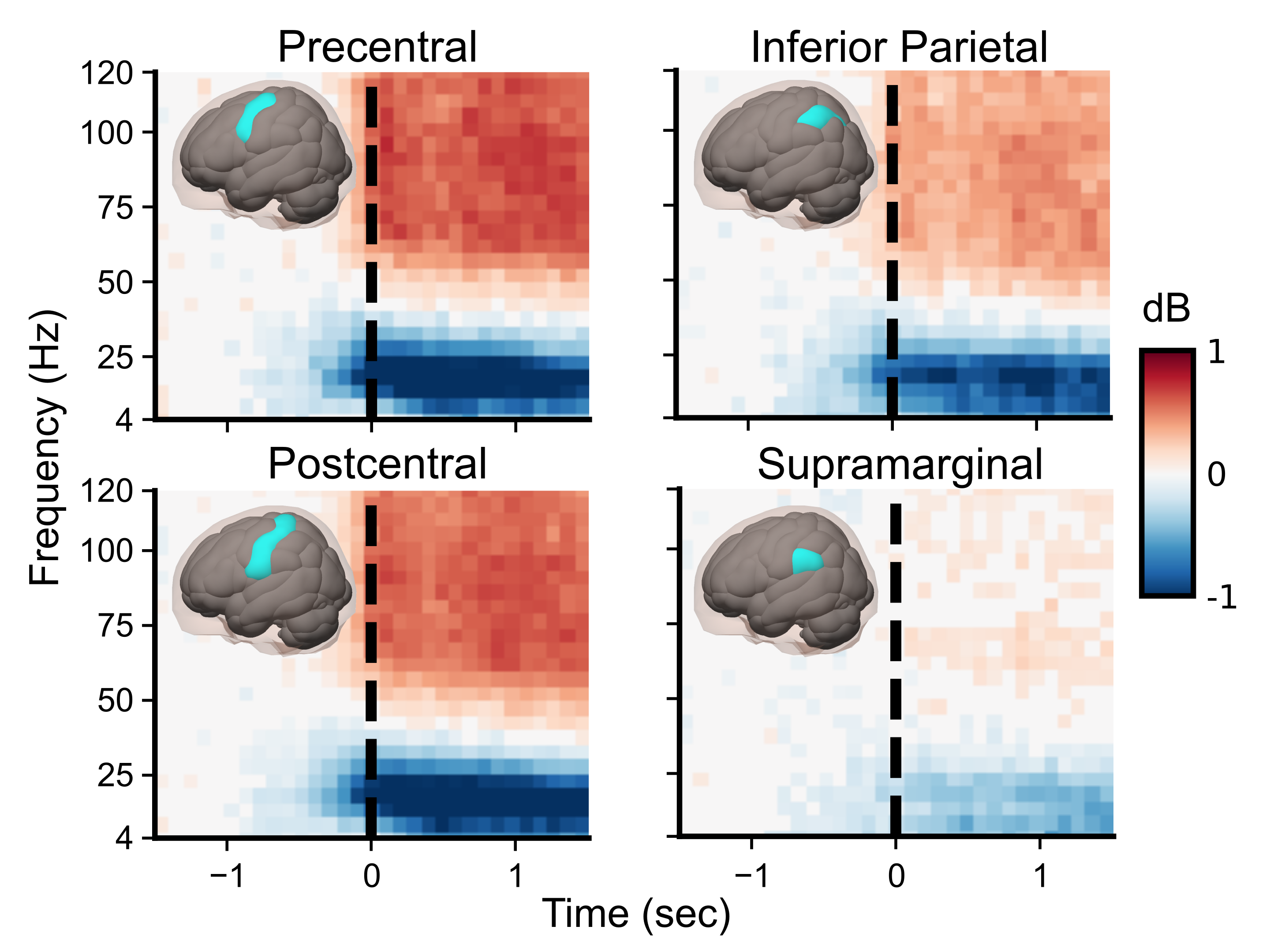}
    \caption{ 
    Neural correlates of movement initiation:
    Event-locked spectrograms, averaged by brain region (cyan color in insets) across 12 subjects, showed movement-associated high-frequency power increase and low-frequency power decrease.
    These patterns corroborate and strengthen previous findings from controlled experiments \protect\citep{miller_spectral_2007}.
    See our companion preprint \protect\citet{peterson2020behavioral} for a deeper exploration of the behavioral and neural variability of these movements.}
    \label{fig_spectrograms}
\end{figure}

\subsection{Neural correlates of movement initiation}
We observed movement-associated power increases in a high-frequency band (76--100 Hz) and decreases in a low-frequency band (8--32 Hz) across many cortical areas (Figure~\ref{fig_spectrograms}).
This pattern is similar to what has been observed with controlled experimental trials in \citet{miller_spectral_2007}, \citet{volkova2019decoding}, and \citet{Yuan2014BrainComputerIU}.
Furthermore, we strengthen prior findings by showing that these movement-associated patterns hold across 5 consecutive days, well beyond the timescale of a typical controlled experiment (less than hours).
To the best of our knowledge, this is the first reported instance of a TF analysis of spontaneous naturalistic movements using events discovered by an automated workflow. 
We elaborate on these results in our concurrently released preprint \citep{peterson2020behavioral}, where we further investigated the consequences of the relatively higher variance of naturalistic movement statistics and modeled the contributions of the various movement metadata to the observed neural responses.

\subsection{Decoding naturalistic movement initiation}
\begin{figure}[tb!]
    \centering
    \includegraphics[width=1.0\linewidth]{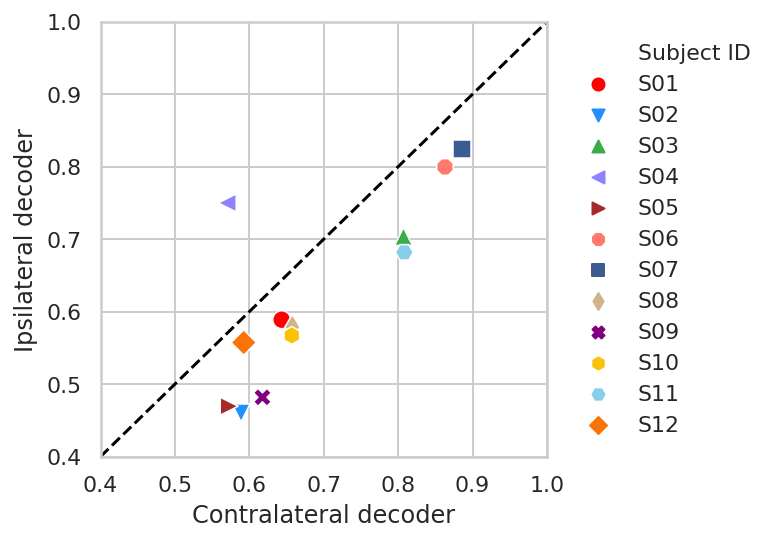}
    \caption{Test set decoding accuracy for initiation of movement of contralateral (side opposite electrode implant) and ipsilateral (same side) wrists:  
    As expected, decoding of contralateral movements is slightly more accurate than ipsilateral in almost all cases.
    }
    \label{fig_accuracy}
\end{figure}

\begin{figure*}[tbhp!]
    \centering
    \includegraphics[width=1.0\linewidth]{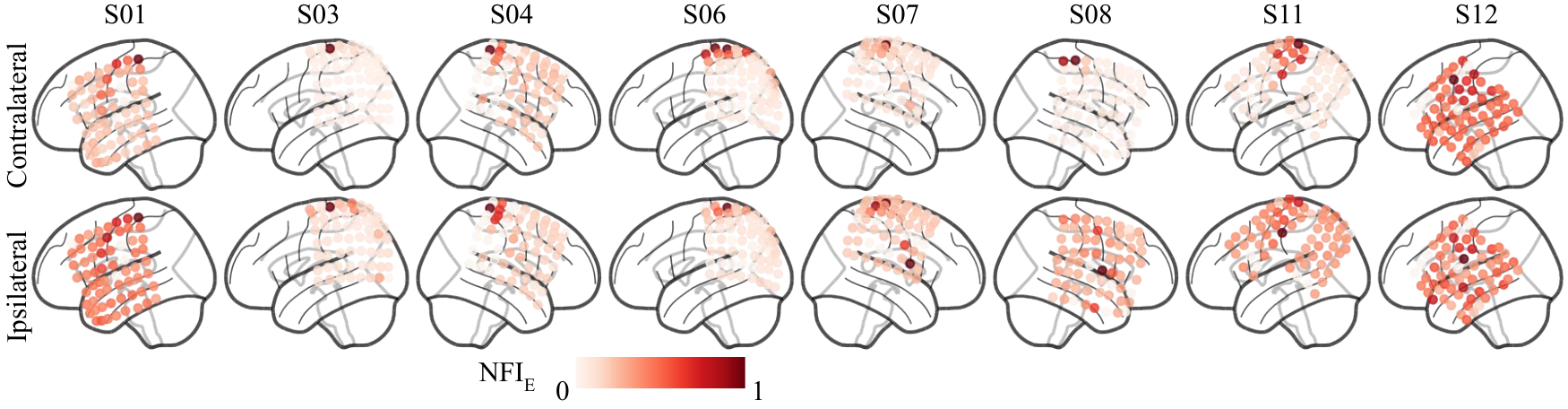}
    \caption{Contralateral and ipsilateral wrist movement initiation decoder feature importance scores aggregated by electrode (NFI$_E$), 
    showing spatial contributions of different brain regions. 
    Scores are normalized by dividing by highest electrode score for each decoder. 
    Electrode coverage over motor cortex is highly correlated with decoder accuracy; 
    for instance, subjects having good motor cortex coverage (S07, S06, S03 and S11) have the highest decoding performance (Figure~\ref{fig_accuracy}).
    See Figure A\ref{fig_nfi_e12} for plot with all 12 subjects.     
    }
    \label{fig_nfi_e8}
\end{figure*}

\begin{figure}[tbhp!]
    \centering
    \includegraphics[width=0.99\linewidth]{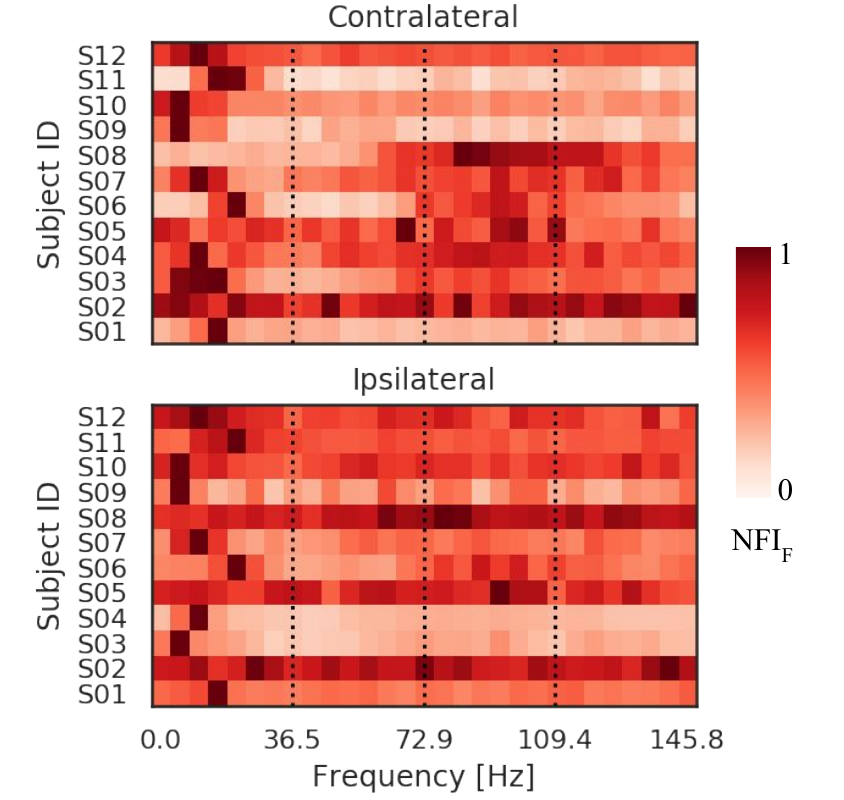}
    \caption{
    {
    Decoder feature importance scores aggregated by frequency and normalized by dividing by score of highest frequency bin per subject (NFI$_F$). 
    Heatmaps show that the most relevant spectral features tend to come from a low-frequency band ($<$ 35Hz) and a high-frequency band (around 100 Hz), similar to prior findings \protect\citep{miller_spectral_2007}. 
    When motor cortex electrode coverage is lacking (e.g. contralateral S02 and S05) or if ipsilateral wrist movement is being decoded, spectral feature contributions tend to be more broadly distributed across the frequency spectrum, and correlated with lower decoding accuracy.
    } 
    }
    \label{fig_nfi_freq}
\end{figure}

Individual subject classifier performance varied widely between subjects, ranging from around chance levels to $80\%$ on test accuracy (Figure~\ref{fig_accuracy}), comparable to previously reported work \citep{gabriel_neural_2019-1,wang2018ajile}.
Classifier performance tended to be correlated with extent of motor cortex coverage (Figure~\ref{fig_nfi_e8} and A\ref{fig_nfi_e12}).
Due to hemispheric lateralization of brain function, decoding contralateral limb movements is expected to be more accurate than decoding ipsilateral movements \citep{tam2019human}. 
Consequently, test set decoding accuracy (Figure~\ref{fig_accuracy}) was higher for the contralateral wrist in almost all subjects. 
Since false positives (FPs) in the event data establish a ceiling on classifier accuracy, we estimated their prevalence by manually inspecting 100 randomly sampled events per event-type from each subject 
(5$\%$ $\pm$ 10$\%$ s.d. for no-movement, 22$\%$ $\pm$ 16$\%$ s.d. for contralateral, and 14$\%$ $\pm$ 10$\%$ s.d. for ipsilateral events). 
With more stringent rejection of FPs, we expect improved decoding performance and potentially more pronounced differences between contralateral and ipsilateral decoding accuracy.  

{
To interpret the importance of spectral features in the decoder, we  visualized Random Forest feature importance scores  \citep{Breiman:2001:RF:570181.570182, hastie2009elements}. 
We aggregated these scores in two ways to gain insight into their spatial (Figure~\ref{fig_nfi_e8} and A\ref{fig_nfi_e12}) and frequency (Figure~\ref{fig_nfi_freq}) components. 
Spectral features are indexed by electrode, time, and frequency. 
We define Feature Importance aggregated by Electrode $FI_E(e)$ for electrode $e \in \mathcal{E}$ as:
$${FI}_E(e) = \sum\limits_{ f \in \mathcal{F}, t \in \mathcal{T} } FI(t, f, e),$$
where $\mathcal{E}, \mathcal{F}$ and $\mathcal{T}$ are the sets of electrodes, frequency-bins, and time-bins over which the spectral features are calculated, respectively. 
For the purpose of visualization, we normalized these values to get Normalized Feature Importance aggregated by Electrode $NFI_E(e)$:
$$NFI_E(e) = \frac{ FI_E(e) }{ \max_{e \in \mathcal{E} } FI_E(e) }. $$
As seen in Figures \ref{fig_nfi_e8} and A\ref{fig_nfi_e12}, we found that electrodes over sensorimotor cortex, when available, dominated feature importance (e.g. Subjects S07, S06, S03 and S11 in Figure~\ref{fig_accuracy}). 
When motor cortex coverage is limited or unavailable, decoding is still possible because inter-region correlations are known to exist in the brain \citep{tam2019human,miller_spectral_2007,schalk2007decoding} and were likely exploited by the classifier, but with limited decoding capacity.

To understand the contributions of various frequency-bins to decoding, we define analogous formulas for feature importances aggregated by frequency-bin:     

$$FI_F(f) = \sum\limits_{ e \in \mathcal{E}, t \in \mathcal{T} } FI(t, f, e)$$

$$NFI_F(f) = \frac{ FI_F(f) }{ \max_{f \in \mathcal{F} } FI_F(f)}$$

As seen in Figure~\ref{fig_nfi_freq}, we found that a low-frequency band ($<$ 35Hz) and a high-frequency band (around 100 Hz) dominates feature importance, similar to prior findings \citep{miller_spectral_2007}.   
If ipsilateral wrist movement was being decoded, or when motor cortex electrode coverage was lacking (e.g. contralateral S02 and S05 in Figures \ref{fig_nfi_e8} and \ref{fig_nfi_e12}), spectral feature contributions tended to be more broadly distributed across the frequency spectrum and correlated with lower decoding accuracy.
    
\section{Discussion}

In summary, we have developed a highly automated and scalable approach for analyzing long-term datasets of simultaneously collected human brain and naturalistic behavior data.
Our workflow robustly uncovered and annotated thousands of human upper-limb movement events in behavior videos. 
To detect movement events, we first discretized pose time-series for each wrist into two latent states, indicating movement or rest, and then used regular expressions to look for user-specified patterns in the latent state sequences. 
This semi-supervised strategy allowed us to rapidly explore movements and their associated brain responses.
Importantly, our curated naturalistic dataset supported direct comparison with existing literature from controlled experiments.
To demonstrate the applicability of our workflow, we analyzed the brain data associated with the annotated events from two perspectives:
characterizing neural correlates of movement, and decoding naturalistic movement initiation using ECoG data.
Key to the success of our applications is the availability of a large number of repeated instances of movement initiation events, all available with high temporal precision, which is an essential requirement for generating event-averaged spectrograms \citep{cohen_analyzing_2014}.  
The ability to select movements by magnitude, onset velocity, and complexity (using shape metadata) allowed us to match movement statistics between naturalistic and controlled experimental data, enabling a fair comparison.
Furthermore, the ability to select events without opposing wrist activity allowed us to disambiguate confounds when comparing movement decoders for opposing wrists.     

\subsection{Limitations and future work}
Our work has a number of limitations that can be improved with further development.
First, our strategy of discretizing individual keypoint time-series to two latent states and then pattern-matching on latent state sequences may be challenging with more complex behaviors involving coordinated movement of more keypoints. 
When we increased the number of latent states in the pose segmentation process, we also noticed that behavioral states were harder to interpret and associated ECoG responses were not easily separable.
The automated analysis of behavior for simple model organisms such as worms \citep{gupta2019context}, zebrafish \citep{johnson2020probabilistic}, flies \citep{berman2016predictability} and mice \citep{luxem2020identifying, markowitz_striatum_2018,datta2019q}, has advanced to the extent of being able to automatically extract hierarchies of coordinated behavioral sequences (or \textit{grammars}) from naturalistic videos.
Except for some very limited work \citep{SummersStay2012UsingAM, Yang2013ACS}, such progress has been elusive in human computer vision, possibly due to the sheer complexity and variability of human movements in various contexts.
Though not tailored to our temporal precision requirements, future research in fine-grained human action recognition in sports \citep{shao2020finegym, piergiovanni2018fine}, domestic \citep{rohrbach2012database} and industrial \citep{kobayashi2019fine} contexts could eventually provide methods that enable the collection of massive datasets of finely annotated human behavior.


All of our data was acquired opportunistically and videos were recorded from a single clinical monitoring camera.
Thus, a primary drawback of the event metadata generated by our pipeline is that they are derived from pose-estimation on single-camera RGB images, implying that all pose coordinates are 2D projections and that the fidelity of pose-derived metadata is limited. 
However, one can still extract utility from the event metadata by coarse-graining or binning them.
The kinematic dataset could be made significantly richer by the use of additional hardware, such as an RGB-D (RGB with depth) camera or a stereoscopic camera system, that would enable pose-estimation and object-tracking in 3D \citep{karashchuk2020anipose, hansen2019fusing, sarafianos20163d}. 

We controlled false positives in the event discovery process using a combination of pose-estimation confidence and a tedious manual omit-listing process.
We found the confidence estimate provided by our pose-estimation tool to  perform well under conditions of good visibility, but it was sensitive to variations arising from naturalistic lighting and occlusions.
One potential source of improvement could come from using pose-estimation algorithms that employ body models, such as OpenPose \citep{cao2018openpose}.
In our assessment, DeepLabCut \citep{mathis2020deep, nath2019using} offered a better speed (cost) vs. accuracy tradeoff at the scale we deployed for pose-estimation.
Future work is poised to take advantage of rapid innovations in computer vision, as more tools become available and accessible.
While manual creation and review of an omit-list cannot be completely avoided for compliance with human research protocols, we believe that a stereoscopic or depth based camera system could also help detect occlusions better and lead to a reduction in false positives.

Finally, two limitations arise from the opportunistic data-collection paradigm itself.
First, we have limited our study to a subject's wrists because they are relatively unconstrained and can perform spontaneous naturalistic movements compared to the rest of the subject's body.  
Our subjects' heads are tethered to a brain recording device that partially restricts the movement of the rest of their upper body.
The study of more naturalistic, especially more active, behaviors would require wireless recording.
Second, ECoG data such as ours has been obtained opportunistically from a neuro-atypical patient population undergoing long-term monitoring preceding invasive epilepsy resection surgery. 
We note with caution that conclusions from analyzing such data might not generalize well to the broader, neuro-typical population.

\subsection{Opportunities using curated dataset}
Accompanying our manuscript, we have publicly released our curated dataset comprising neural data and event metadata for over 40,000 instances of naturalistic human upper-limb movement events, and an equal number of rest events, across 12 subjects over about a week of clinical monitoring each.
We expect our dataset to be broadly applicable to BCI research like previously released datasets such as the BCI competitions I--IV datasets \citep{sajda2003data, blankertz2004bci, blankertz2006bci, tangermann2012review}, or other ECoG data libraries \citep{miller2019library} that were generated through controlled experimentation.
While the aforementioned datasets consist of 10s--100s of repeated instances of a behavior per subject, our dataset provides 1000s of instances per subject.
It captures rich naturalistic variability across multiple axes relating to the neural activity (subject, seizure foci, day of observation, electrode placement, and recording fidelity) and the behavior (subject activity profile, medication and treatment regime; wrist movement times, movement handedness and sequencing; and other event metadata). 

We are working on refinements and extensions of the preliminary investigations presented here, and believe that our dataset could serve several other lines of scientific inquiry. 
A thorough analysis of neural encoding of wrist movement initiation is available in our simultaneously released preprint \citep{peterson2020behavioral}. 
As a follow-up to the limited prototype described here, we are currently exploring the use of neural networks to build more expressive decoders \citep{roy2019deep} for events and their associated event metadata.
The abundant availability of neural data also allows us to explore representation learning to obtain interpretable task-specific neural features \citep{pailla2019autoencoders, shiraishi2020neural}, and transfer learning to adapt decoders trained for one subject to another \citep{wu2020transfer, elango2017sequence, shenoy2007generalized}.

Our dataset is amenable to several types of modeling goals and approaches including
unsupervised latent factor modeling to extract single-trial neural dynamics \citep{pandarinath2018latent, ly2018electrocorticographic, cole2019cycle, pandarinath2018inferring, zhao2016interpretable}, 
dynamical modeling of the electrocorticographic spectrum \citep{chaudhuri2018random, beck2018state, haller2018parameterizing, brunton2016extracting},
probabilistic modeling to better understand neural data variability across trials, subjects and brain-regions \citep{omigbodun2016hidden, yang2019dynamic, abbaspourazad2018identifying, yang2017dynamic},
generative modeling to generate synthetic brain data \citep{hartmann2018eeg, aznan2019simulating}
and modeling the non-stationarity of the brain signal over long recording time spans \citep{farshchian2018adversarial, klosterman2016day, shenoy2006towards}. 
We hope that our dataset will enable further research on models of neural function that incorporate naturalistic variability.

The presence of false positives in the data also motivates exploring algorithms for machine learning with noisy labels \citep{rolnick2017deep, natarajan2013learning, han2018co}. 
This paradigm has been well studied for other applications of machine learning such as computer vision \citep{li2017learning}, where amassing large datasets with noisy labels is relatively inexpensive, but quality labeling is expensive to obtain. 
The large behavioral variability associated with our neural data could also be used to investigate optimal training set selection, i.e. what types of and how much training data could be ideal for training a decoder \citep{wei2015submodularity, wei2014unsupervised, krause2008robust}. 
Such characterizations could be used to inform the engineering of BCIs, making them significantly more robust to the variations present in real-world deployments.

\section*{Code and dataset release}
Code to reproduce several key plots in this manuscript is publicly available at: \url{https://github.com/BruntonUWBio/singh2020}. 
Our complete curated dataset consisting of events and their metadata and associated neural data can also be downloaded following the instructions provided at the aforementioned URL.

\section*{Acknowledgements}

We thank John So for extensive help with manual annotation of the video data.
This work benefited from and was enabled by the groundwork laid by Nancy X. R. Wang towards study approval, initial clinical data procurement, preprocessing, and manual annotation, and establishing the plausibility of movement initiation prediction using a subset of this clinical data.  
We thank the neurosurgeons Dr. Jeffrey G. Ojemann and Dr. Andrew Ko, and the staff and consenting patients at the University of Washington Harborview Medical Center in Seattle, for supporting this research.
We thank Pierre Karashchuk, Kameron D. Harris, James Wu, Nile Wilson, Ariel Rokem, Renshu Gu, David J. Caldwell, and Preston Jiang for helpful discussions and suggestions.
This work was funded by NSF award (1630178) and DOD/DARPA award (FA8750-18-2-0259) to BWB and RPNR, NSF award EEC-1028725 to RPNR, the Alfred P. Sloan Foundation (BWB), and the Washington Research Foundation (BWB).

\section*{Author Contributions}
SHS, RPNR, and BWB conceived of the study/analysis. 
SHS and SMP performed the data analysis. 
SHS, SMP, RPNR, and BWB interpreted the results. 
SHS and BWB wrote the manuscript. 
SHS, SMP, RPNR and BWB edited the manuscript. 
RPNR and BWB acquired funding for the project.

} 

\newcommand{\newblock}{} 
\bibliographystyle{abbrvnamed}
\interlinepenalty=10000 
\bibliography{jne2020}

\appendix
\begin{figure*} 
    \centering
    \textbf{APPENDIX}
    \includegraphics[width=1.0\linewidth]{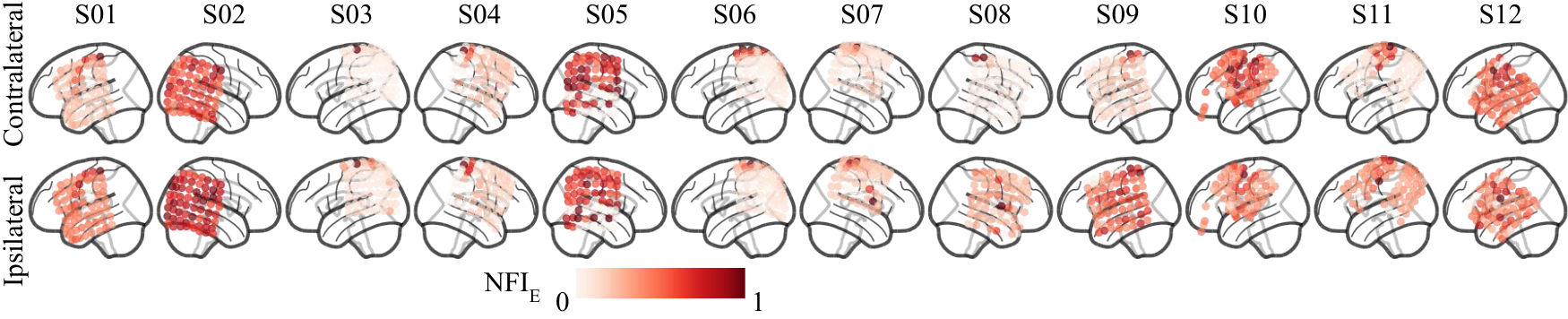}
    \caption{Contralateral and ipsilateral wrist movement initiation decoder normalized feature importance scores aggregated by electrode (NFI$_E$), 
    showing spatial contributions of different brain regions for all 12 subjects. 
    We see the same trend of motor cortex coverage being correlated with decoding accuracy, as was noted in Figure~\ref{fig_nfi_e8}. 
    Subjects having good motor cortex coverage (S07, S06, S03 and S11) have the highest decoding performance (Figure~\ref{fig_accuracy}).
    Additionally, we see that electrodes with high normalized feature importance tend to be more spatially localized in the case where good motor cortex coverage is available. 
    }
    \label{fig_nfi_e12}
\end{figure*}

\begin{figure*} 
    \centering
    \includegraphics[width=1.0\linewidth]{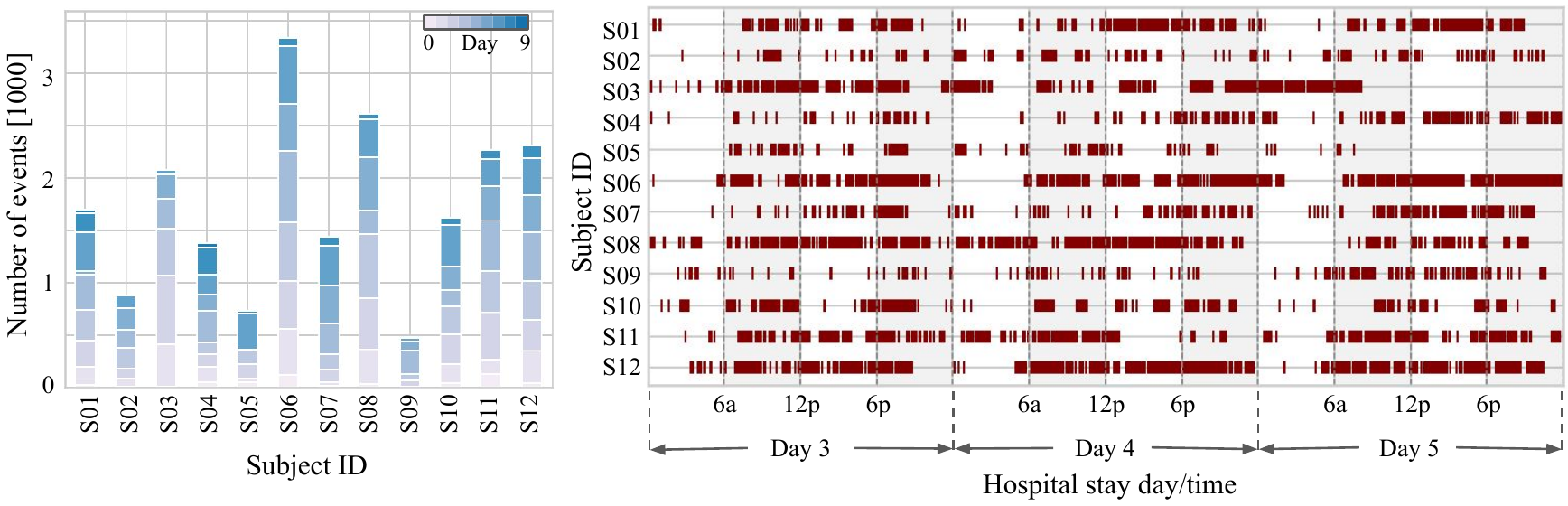}
    \caption{ 
    Number of left-wrist movement initiation events discovered per day for each of 12 subjects, totaling 484 to 3338 events per subject across their entire duration of clinical observation 
    (219 $\pm$ 104 s.d. per day).
    [Right] Raster plot of left-wrist movement initiation occurrences.  
    See Figure~\ref{fig_rastercounts_r} for equivalent plots for the right-wrist. 
    }
    \label{fig_rastercounts_l}
\end{figure*}

\begin{figure*} 
    \centering
    \includegraphics[width=1.0\linewidth]{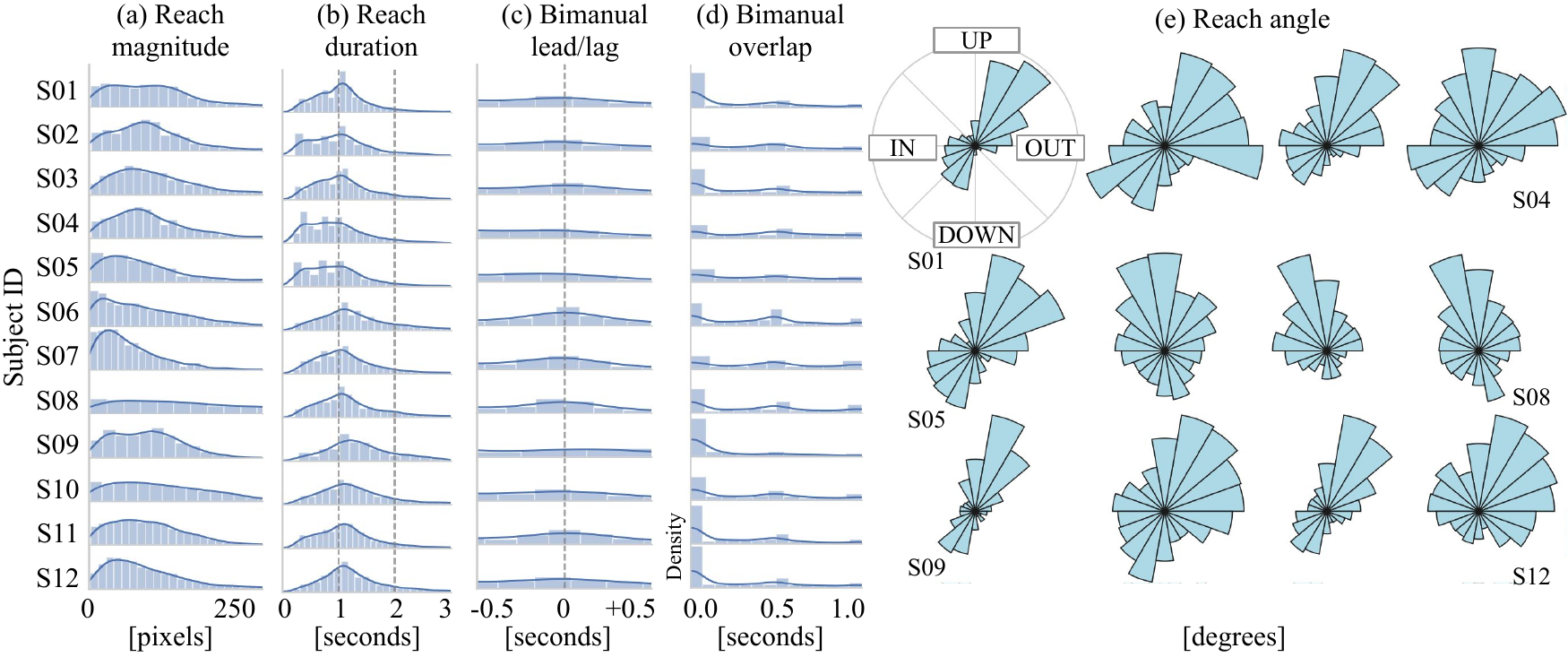}
    \caption{ 
    Histograms of left-wrist movement initiation event metadata per subject for their entire duration of clinical observation. 
    We see all the same trends for the left-wrist as were noted for the right-wrist in Figure~\ref{fig_metadata_r}.
    }
    \label{fig_metadata_l}
\end{figure*}

\end{document}